# Confinement of electron holes via the peroxo group formation in the negative charge-transfer materials on the example of SrFeO$_3$: plane-wave density functional theory predictions

Nikita A. Afimchenko, Aleksandr A. Shubin, Igor L. Zilberberg[*], Alexander P. Nemudry

Institute of Solid State Chemistry and Mechanochemistry, Novosibirsk 630090, Russia

**Abstract**

The present work puts forward a concept that the thermostable O1s XPS peaks with energy of about 531 eV in negative charge-transfer SrFeO$_{3-\delta}$ perovskite are determined by the peroxo-like oxygen species. The peroxo group forms via coupling two oxygen anions coordinated to iron cations with d$^5\underline{L}$ ($\underline{L}$-oxygen electron hole) configuration. By means of plane-wave DFT+$U$ approach there have been shown that the peroxo group represents a metastable state in the absence of oxygen vacancies nearby. The O-O bonding confines two electron holes freezing the "3+" oxidation state for two iron cations bridged by peroxide. Increasing the peroxo group numbers makes the ferrite a semiconductor with charge-transfer gap of about 0.6 eV.

[*] Corresponding authors: zilberberg@solid.nsc.ru

# Introduction

The formation of peroxo group in the $ABO_3$ perovskites and perovskite-like oxides has been theoretically considered in connection with the concept of the negative charge-transfer materials having strong mixing between $d^n$ and $d^{n+1}\underline{L}$ configuration of the B cation with electron hole $\underline{L}$ in (usually oxo) ligand [1–3], and the B-O covalency [4,5] as well as one of Frenkel defects [6]. In addition the O-O peroxo-like dimers in high-capacity layered oxides for Li-ion batteries were visualized via transmission electron microscopy and neutron diffraction in parallel with the DFT calculations [7].

The X-ray photo-electron spectroscopy (XPS) evidences on multiple oxygen electron states in perovskites $ABO_{3-\delta}$ as seen from O1s XPS for $LaCo_{1-x}Fe_xO_{3-\delta}$.[8]. In particular, for $La_{1.0}Co_{0.96}Fe_{0.003}O_{3-\delta}$ and $La_{0.89}Co_{0.94}Fe_{0.17}O_{3-\delta}$ the calcination decreases the number of peaks from six (for fresh sample) to four and two peaks at temperatures of 200°C, 400°C and 900°C, respectively, (see Figure 8 in ref. [8]). The authors have concluded that there are two distinct lattice states of oxygen with energies of 529.2 and 531.7 eV which present up to 900°C. The low binding energy peak is ascribed to regular lattice oxygen $O^{2-}$, while the 531.7 peak is associated with the $O^-$ form [8].

Analogous O1s states with energies 528-529, 530.8-531.5 eV and 532-533 eV have been detected in the O1s XPS spectra for perovskites $La_{1-x}Sr_xFeO_{3-\delta}$. The lowest-energy peak of 528.3 eV was assigned to lattice oxygen of $O^{2-}$ type, the 531.2 eV peak - to weakly bound atomic oxygen of the $O^-$ type, while highest energy peak of 532-533 eV – to water molecule adsorbed on the surface [9].

Upon heating under UHV up to 500°C the peak of 532-533 eV almost disappeared while the peak of 531.2 eV only looses intensity and the peak of 528.3 eV remains unchanged. In the $O_2$ atmosphere ($P(O_2)$=300 mbar, 20 minutes at 400°C) the intensity of the 531.2 peak comes to about half of its initial value. Repeated ultra high vacuum (UHV) heating up to 500° C leads again to decreasing of the 531.2 peak intensity. The authors of this work explained such a behavior by the presence of two distinct oxo species in the perovskite – so-called strongly bound (528.3) and weakly bound (531.2) atomic oxygen. The referral to mobility in the name of these states is obviously connected with XPS detected behavior of spectra intensity under heating. The oxidation state for these oxo forms is assigned to be 2- and 1-, respectively.

Another example of high-energy oxygen state appears for the $Pr_2NiO_{4+\delta}$ oxide with Ruddlesden-Popper (RP) structure [10]. This oxide also reveals three types of peaks with energies being close to those given above: 529.1, 530.8, and 532.0 eV. Analogously to the case of $La_{1-x}Sr_xFeO_{3-\delta}$ there was applied the procedure of heating of $Pr_2NiO_{4+\delta}$ in UHV up to 500°C followed by cooling down to room temperature and heating in $O_2$ atmosphere (with pressure of 300 mbar) at 350°C to see how the peak intensity changes with the entry of oxygen into material.

The appearance of high energy O1s peaks is not restricted only to perovskite-like oxides. In particular, the peak of 531.2 eV has been detected in O1s XPS spectra for CuO films [11]. The authors of this work had assigned this peak to associative form of oxygen.

As one can see there are quite similar oxygen forms in different perovskite and perovskite-like systems. The minimal energy peaks are assigned to lattice monoatomic oxygen form of the $O^{2-}$ character. The maximal energy peaks at about 532 eV are usually assigned as surface molecular species ($O_2^-$ and $H_2O$) which are proved to desorb from the surface at 400-500 K. The most intriguing oxygen state is associated with the thermostable peaks around 531 eV. The assignment of this state is quite controversial.

The present work assigns the XPS peaks with energy of about 531 eV in $SrFeO_{3-\delta}$ and analogous peaks in other perovskites and RP oxides as determined by the peroxo-like oxygen species which form at elevated temperatures in the bulk and on the surface of the oxides. The O-O bonding obviously decreases the electron density on each oxygen atom in peroxo species making its O1s state energy in XPS substantially increased. The intensity drop for the XPS peak of 531 eV under heating up to 500 K for above given oxides can be connected with the migration of peroxo groups to the surface and subsequent desorption in the form of molecular oxygen. To obtain an assignment for various oxygen forms in perovskites there has been used the O2s localized band energy within the standard DFT+$U$ calculations.

## Calculation details

All calculations have been performed with the Vienna Ab Initio Simulation Package (VASP 5.4.4. version) [12,13]. There was applied PW91.6 ( Sr_sv, Fe, O) exchange-correlation functional along with the projector-augmented wave (PAW) approximation for the electron interaction with ionic core [14,15]. The threshold for the electronic self-consistency was chosen to be $10^{-6}$ eV while that for lattice ions relaxation was $10^{-5}$ eV. To sample Brillouin zone the 4x4x4 mesh of the k points centered at Gamma point was used. The energy cutoff for the plane-wave basis set was chosen to be 600 eV. To cure underestimation of band gaps in DFT the DFT+$U$ approach (to account for the effects of strong correlation) with $U$=3.9 eV has been applied [16] [17]. Transition states (TS) were determined by means of the NEB in conjunction with the improved dimer method [18,19]. To be sure that found TS corresponds to proper reactants and products IRC method was used.

To assign high-energy O1s states in XPS there has been applied a simple approach in which the information about the O1s energy shift (due to changeable electron density on oxygen) is obtained from the shift of the O2s state. Unlike the O1s state being absent in the standard plane-wave approach due to the pseudopotential substituting core atomic states, the O2s state is present in the calculation. A change in the electron density on oxygen site in particular environment must cause a shift of the 1s(O) and 2s(O) energies coherently: both states shift down in case of smaller electron density and up in case larger electron density. These theoretical shifts in case of low and high electron density at oxygen site correspond to the O1s peaks of higher and lower energy, respectively. In practice to distinguish between various oxygen states one can use oxygen atom partial density of states. This approach was first applied by one of us in [20].

## Perovskite model

The $SrFeO_3$ was calculated for stoichiometric composition with $\delta=0$ for the 2x2x2 cell. So-constructed models consist of regular lattice oxygens, oxygen associates (peroxo-like groups) and oxygen vacancies formed due to the oxygen association (one vacancy for each peroxo group) or the migration of regular oxygen into the interstitial position. Oxygen vacancies form in fact as a result of Frenkel defects. In Figure 5 there are plotted the model with single peroxo group and single interstitial oxygen. The four peroxo groups model will be examined too as an example of maximal number of peroxo groups for stoichiometric strontium ferrite where the number of electron holes on oxygen are equal to the number of iron cations $n_{Fe}$ and so the number of peroxo groups is $1/2n_{Fe}$. For the latter model each iron cation has to possess the $d^5$ configuration.

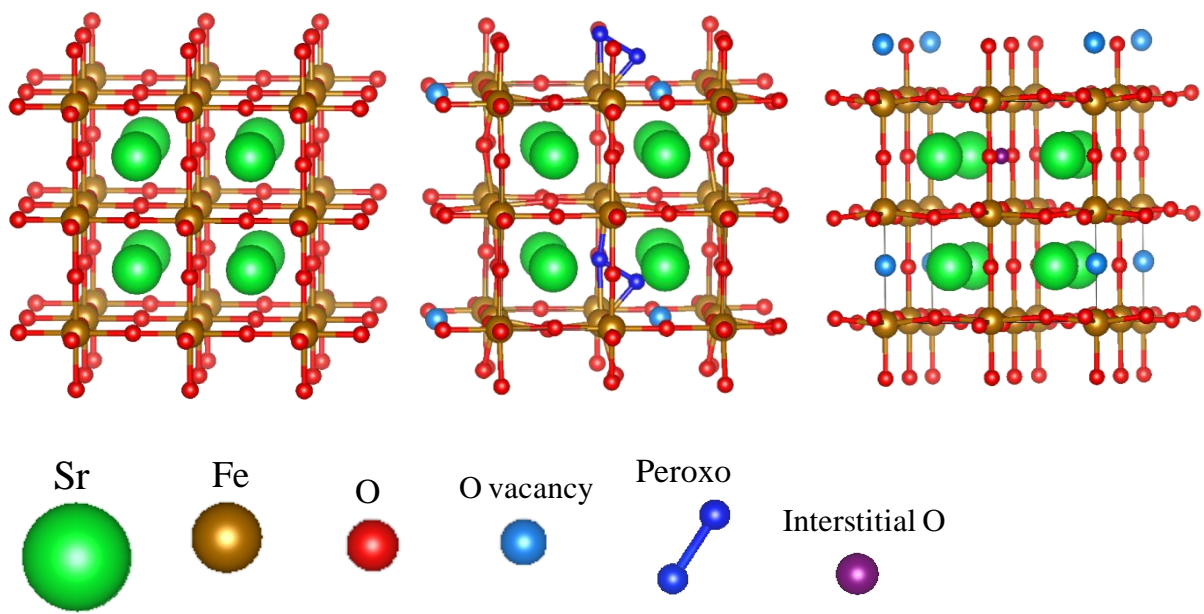

Figure 1. The 2x2x2 $Sr_8Fe_8O_{24}$ supercell for simulating regular oxo site (left), $Sr_8Fe_8(O-O)O_{22}$ perovskite with single peroxo group and corresponding single oxygen vacancy (middle) and the interstitial oxygen (right).

## Results

*Electron structure for various oxygen sites*

The partial density of the O2s states for above described perovskite models shows that the O2s energy for peroxo-like species and interstitial site substantially deviates from the regular lattice oxygen (Figure 2). Without peroxo group 2s oxygen states are broaden reflecting admixture of the Sr and Fe core states. The peroxo group brings into the partial density of states (PDOS) a new localized state lying below regular lattice 2s(O) states by about 4 eV. This state seems to fingerprint the O-O bonding between 2s states. Thus, the oxygen in peroxo group (having formal charge of "-2") becomes less negatively charged than the regular lattice oxygen. This corresponds to the change of formal oxidation state from "2-" for regular lattice oxygen to "1-" for oxygen in peroxo group. The difference in the XPS O1s energies (528.3 and 531.2 eV) is about 3 eV being well agreed with found calculated difference. Interpreting the latter result one should take into account that valence O2s state is certainly more sensitive to the charge of oxygen than core O1s state and, therefore, the variation of O2s state energy is larger than that for O1s.

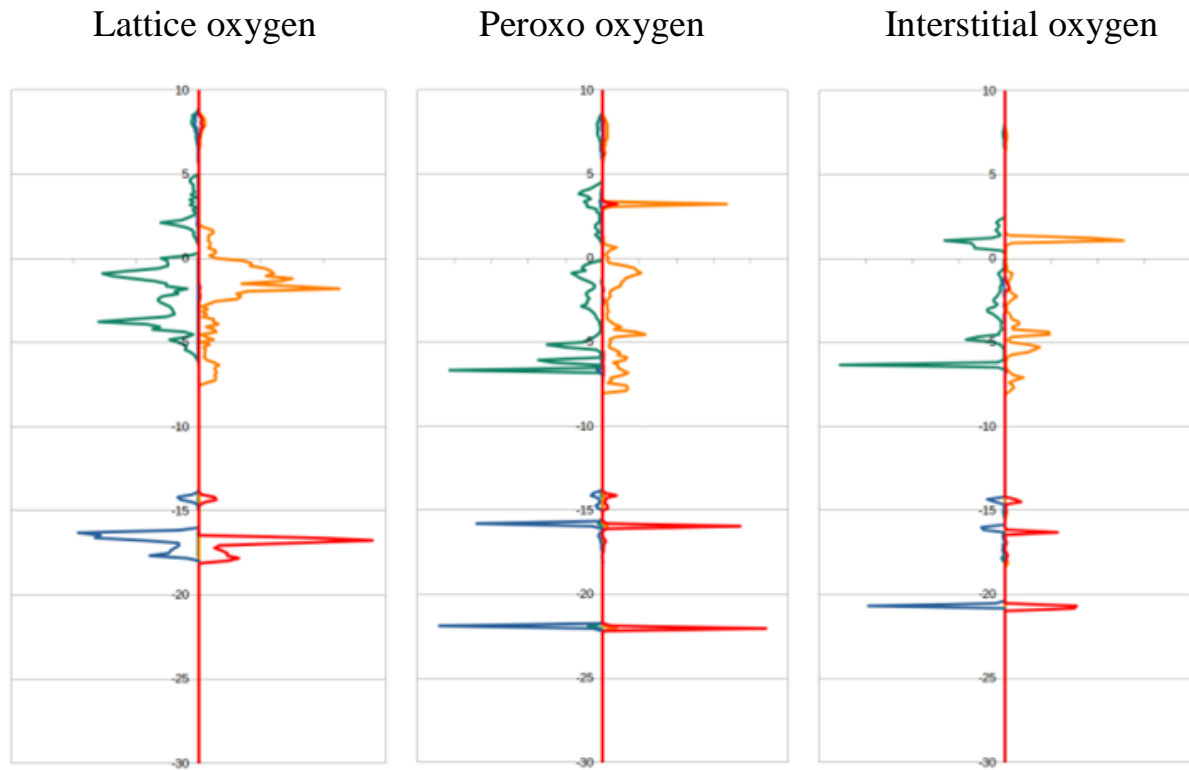

Figure 2. Appearance of additional low lying localized state for peroxo and interstitial Frenkel defect in SrFeO$_3$: partial density of 2s and 2p states for oxygen in regular lattice site (left), peroxo site (middle) and interstitial site (right). Color scheme: p spin up – orange, s spin up – red, p spin down – green, s spin down – blue.

PDOS for interstitial oxygen looks like peroxo oxygen being though even more localized. This reflects the absence of chemical bonding between interstitial oxygen and surrounding metal cations and, correspondingly, the absence of detectable atomic *sp* hybridization. The same arguments explain why the lowest localized pure atomic 2s state for interstitial oxygen (downshifted by about 2.5 eV with respect to that for lattice oxygen) lying higher than the lowest peroxo state. Curiously enough that the peroxo group in BSCF perovskite considered theoretically by Kuklja et al [6] as one of Frenkel defects was named "split interstitial", although the authors did not mention the similarity of the electron structures for these defects.

Worthwhile noting that the oxygen association only slightly deforms the cubic symmetry of the ferrite and lattice parameters. Single peroxo group increases these parameters by less than 0.1 Å (Table 1) though the increase of lattice constants for the ferrite with maximal number of peroxo groups achieves about 0.2 Å.

Table 1. The $Sr_8Fe_8O_{24}$ perovskite calculated in regular cubic form, with single interstitial oxygen (Frenkel defect), single peroxo group, four peroxo groups: lattice constants, relative energy ($\Delta E$)

| System | Lattice parameters, Å | | | $\mu(Fe)/\mu_B$ | $\Delta E$, eV |
|---|---|---|---|---|---|
| | a | b | c | | |
| $Sr_8Fe_8O_{24}$ | 3.886[1] | 3.886 | 3.886 | 3.7[2] | 0 |
| $Sr_8Fe_8O_{24}$ interstitial | 3.985 | 3.908 | 3.923 | 4.0[3], 3.8[4] | 3.63 |
| $Sr_8Fe_8(O-O)O_{22}$ | 3.928 | 3.910 | 3.931 | 3.5 and 3.8[5] 3.7[6] | 1.24 |
| $Sr_8Fe_8(O-O)_4O_{16}$ | 4.128 | 4.150 | 4.146 | 4.2[7] | 0.6 |

[1] experimental value is 3.868 [9], 3.845[21]
[2] (experimental 2.7±0.4 for $SrFeO_3$ [22])
[3] for Fe close to interstitial O
[4] for Fe far from interstitial O
[5] for iron atoms in the Fe-O-O-Fe fragment
[6] for Fe far from O-O
[7] experimental 4.6±0.2 for $LaFeO_3$ [23]

Quite important question is the oxidation state of the Fe cation which is nominally "4+", although Fe2p XPS supports the $Fe^{3+}$ oxidation state [2] [9]. Calculated magnetic moment for iron cations in the ferrite varies from 3.7 to 4.2 $\mu_B$ in a series $Sr_8Fe_8O_{24}$, $Sr_8Fe_8(O-O)O_{22}$, $Sr_8Fe_8(O-O)_4O_{16}$, respectively. These values appears to be between experimental $\mu(Fe^{4+})$=2.7±0.4 $\mu_B$ (obtained for iron in $SrFeO_3$ at liquid nitrogen is [22]) and $\mu(Fe^{3+})=4.6\pm0.2$ $\mu_B$ (obtained for LaFeO$_3$ [23]) though the Fe$^{4+}$ state seems to be a minor component (as compared with Fe$^{3+}$) in the iron state even for regular strontium ferrite.

The most remarkable fact here is that the association of oxygen anions to form the O-O bond is equivalent to a "chemical" confinement of electron holes pairs which stabilizes the "3+" oxidation state for iron cations. This effect reveals itself not only in the increase of µ(Fe) up to 4.2 $\mu_B$ but also in the appearance of an energy gap of about 0.6 eV (Figure 3c). In the strontium ferrite the electrical conductivity is considered to be determined by the migration of electron holes along the chain Fe$^{4+}$-O$^{2-}$-Fe$^{3+}$ [24]. The peroxo group formation in our model makes the electron holes trapped in the peroxo groups. As a result all iron cations becomes Fe$^{3+}$ that is the electron hole conductivity disappears. This explains the appearance of energy gap for Sr$_8$Fe$_8$(O-O)$_4$O$_{16}$ which appears to be a semiconductor in which the electric conduction is connected with charge transfer from lattice oxygen as well as peroxo to the *d* shell (Figure 3c). From the other side, discussed peroxo contained model explains the presence of only Fe$^{3+}$ state in XPS for strontium ferrite.

(a)     $Sr_8Fe_8O_{24}$     $\mu(Fe)=3.7$

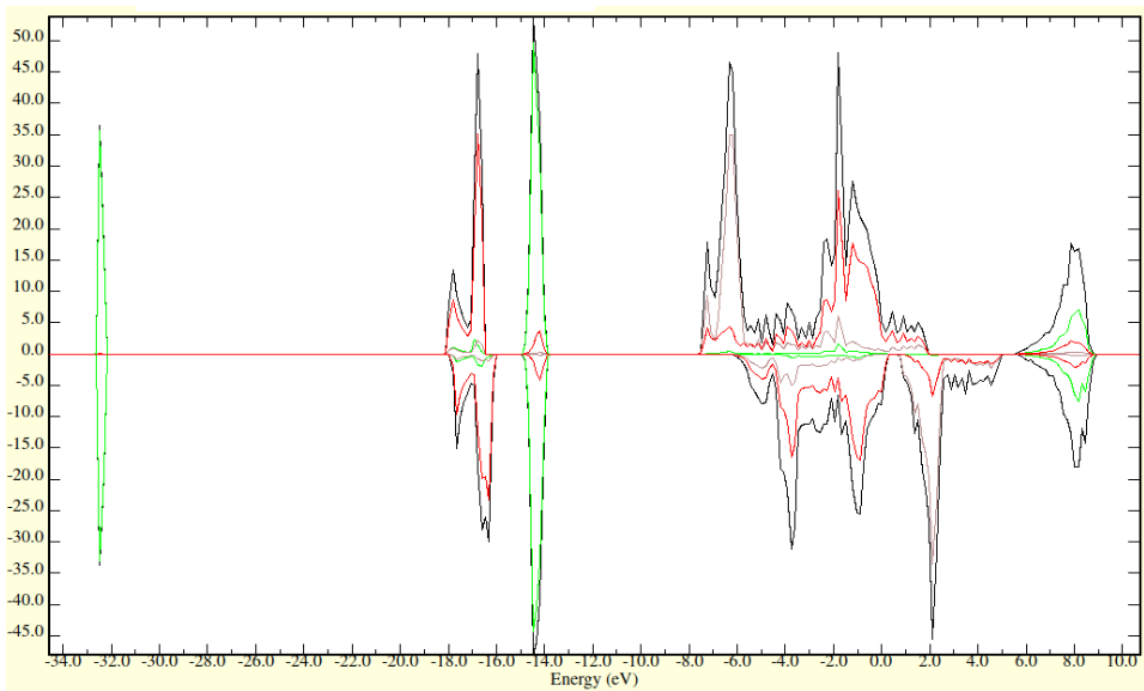

(b)     $Sr_8Fe_8(O\text{-}O)O_{22}$     $\mu(Fe)=3.8$

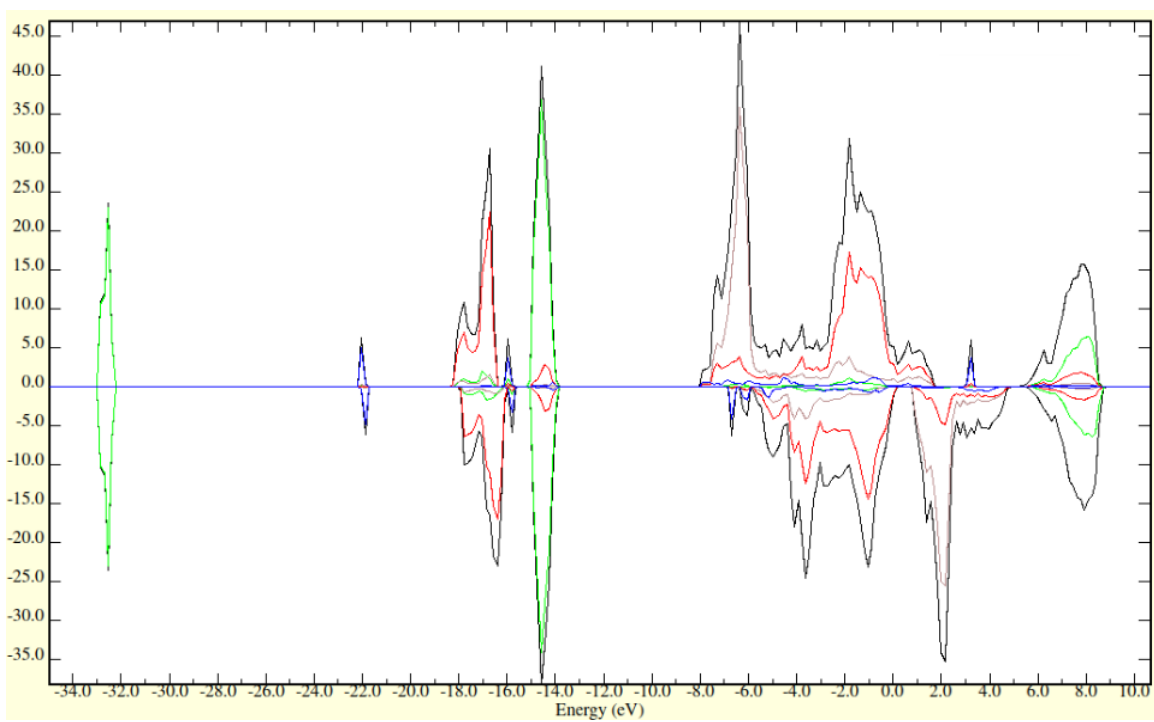

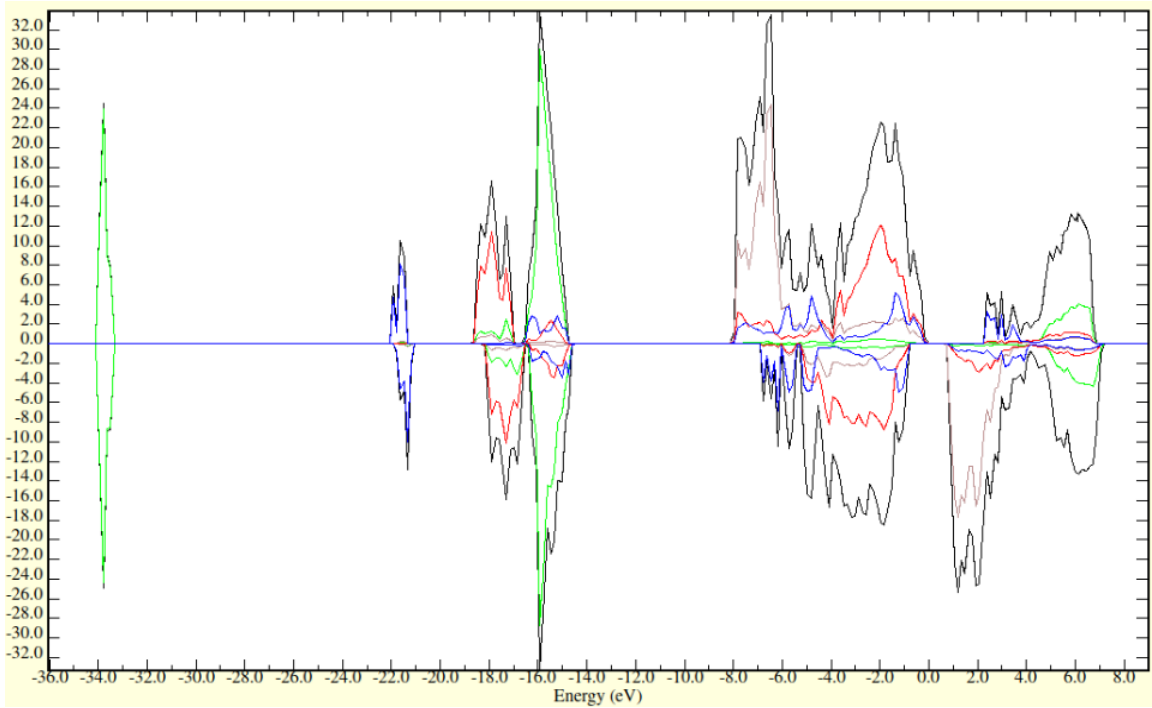

Figure 3. Formation of peroxo groups in the bulk of SrFeO$_3$ - total and partial DOS changes: (a) Regular lattice Sr$_8$Fe$_8$O$_{24}$, (b) Sr$_8$Fe$_8$(O-O)O$_{22}$ single peroxo group, (c) Sr$_8$Fe$_8$(O-O)$_4$O$_{16}$. The top and down parts of DOS correspond to spin up and spin down, respectively. Color scheme: Sr - green, Fe - brown, O - red, O-O (peroxo) – blue.

*Migration barriers*

Migration of regular lattice oxygen is only possible when there is the oxygen vacancy nearby. In its turn the peroxo-like group migration can take place in the absence of oxygen vacancies. It comprises two consecutive acts of the O-O dissociation and the formation of new O-O bond (Figure 4). The barriers have been estimated by means of the NEB and Dimer approaches (Table 1). As one can see the monoatomic oxygen migrates into vacancy over the fairy small barrier of 0.35 eV while the oxygen associate moves over higher barrier of 0.60 eV. The energy difference between the initial and final position for migration has to be zero for lattice oxygen (or vacancy) migration and peroxo group migration since these positions are equivalent by construction. In practice, however these values are small but nonzero perhaps due to ionic relaxation threshold. The energy of peroxo group formation is relatively high (1.25 eV). Nevertheless, taking into account the high-temperature conditions for oxygen transport in perovskite-based device like solid state fuel cell, one cannot ignore a possibility for the peroxo group formation.

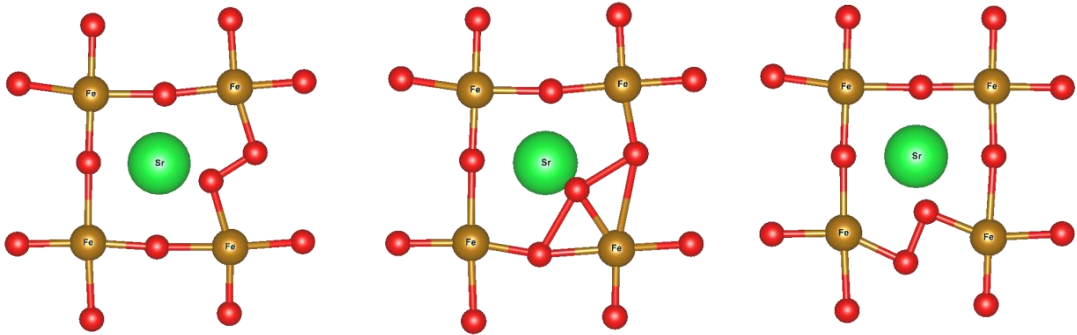

Figure 4. Migration of peroxo group within the $SrFeO_3$.

**Table 2.** Energy difference and barrier (Barrier) for the migration of lattice oxygen into the nearest vacancy (Vacancy migration), the migration of the peroxo-like oxygen group (Peroxo group migration) in the bulk of $SrFeO_3$ lattice as well as the barrier for the peroxo group formation within regular lattice

|  | Vacancy migration | Peroxo group migration | Peroxo group formation |
|---|---|---|---|
| ΔE, eV | 0.11 | 0.02 | 1.25 |
| Barrier, eV, | 0.35 | 0.60 | 1.27 |

## Discussion

The peroxo groups have to form in ferrite membranes within the oxygen transport with relatively high oxygen pressure at the penetration side. This is because the introduction of gaseous oxygen into the surface with negligible number of oxygen vacancies is only possible via the formation of peroxo-like groups. Under such conditions peroxo groups becomes metastable due to the barrier for transferring oxygen atom from one peroxo species to form another.

Although the association of oxygen atoms in the strontium ferrite appears to require a high energy of 1.25 eV, the entropy factor for peroxide (which might be substantial as compared with lattice oxygen taking into account the appearance of vibrational and rotational degrees of freedom) can substantially decrease this energy taking into account the high temperature regime of oxygen transport.

## Conclusion

The 531 eV O1s peaks in XPS for strontium ferrite – known as negative charge-transfer material with the $d^5\underline{L}$ (where $\underline{L}$ – electron hole on oxygen) configuration of iron cation - have been assigned to peroxo-like groups. The latter form from lattice oxygen anions with a calculated barrier of 1.27 eV. The O-O bonding for a pair of nominally $Fe^{4+}$ cations forces both cations to become the $Fe^{3+}$ cations. In the absence of oxygen vacancies around peroxo group it becomes metastable and so the pair of the $Fe^{3+}$ cations coupled with peroxo group. This process becomes a sort of "chemical" confinement for electron holes $\underline{L}$ which are so excluded from metallic conduction via the Fe(4+)- O(2-)- Fe(3+) chain. In addition, the peroxo groups formation makes the ferrite semiconductor with an estimated gap of 0.6 eV being of the charge-transfer in character. The implications of described findings for the insight into the electron structure and various processes like electron and oxygen transport are to be considered elsewhere in details.

## Acknowledgements

The authors acknowledge financial support from the Russian Science Foundation within Project 23-43-00130. The study was carried out using facilities of the Siberian Supercomputer Center at the Institute of Computational Mathematics and

**References**

[1]   J. Zaanen, G.A. Sawatzky, J.W. Allen, Band gaps and electronic structure of transition-metal compounds, Phys. Rev. Lett. 55 (1985) 418–421. https://doi.org/10.1103/PhysRevLett.55.418.

[2]   A.E. Bocquet, A. Fujimori, T. Mizokawa, T. Saitoh, H. Namatame, S. Suga, N. Kimizuka, Y. Takeda, M. Takano, Electronic structure of SrFe4+O3 and related Fe perovskite oxides, Phys. Rev. B. 45 (1992) 1561–1570. https://doi.org/10.1103/PhysRevB.45.1561.

[3]   M.F. Hoedl, C. Ertural, R. Merkle, R. Dronskowski, J. Maier, The Orbital Nature of Electron Holes in BaFeO3and Implications for Defect Chemistry, J. Phys. Chem. C. 126 (2022) 12809–12819. https://doi.org/10.1021/acs.jpcc.2c02545.

[4]   J.B. Goodenough, Perspective on engineering transition-metal oxides, Chem. Mater. 26 (2014) 820–829. https://doi.org/10.1021/cm402063u.

[5]   A. Grimaud, W.T. Hong, Y. Shao-Horn, J.M. Tarascon, Anionic redox processes for electrochemical devices, Nat. Mater. 15 (2016) 121–126. https://doi.org/10.1038/nmat4551.

[6]   M.M. Kuklja, Y.A. Mastrikov, B. Jansang, E.A. Kotomin, The intrinsic defects, disordering, and structural stability of Ba xSr 1-xCo yFe 1-yO 3-δ

perovskite solid solutions, J. Phys. Chem. C. 116 (2012) 18605–18611. https://doi.org/10.1021/jp304055s.

[7] E. McCalla, A.M. Abakumov, M. Saubanère, D. Foix, E.J. Berg, G. Rousse, M.-L. Doublet, D. Gonbeau, P. Novák, G. Van Tendeloo, R. Dominko, J.-M. Tarascon, Visualization of O-O peroxo-like dimers in high-capacity layered oxides for Li-ion batteries, Science (80-. ). 350 (2015) 1516–1521. https://doi.org/10.1126/science.aac8260.

[8] S. Kaliaguine, V. Szabo, A. Van Neste, J.E. Gallot, M. Bassir, R. Muzychuk, Perovskite-type oxides synthesized by reactive grinding, Mater. Sci. Forum. 377 (2001) 39–56. https://doi.org/10.4028/www.scientific.net/msf.377.39.

[9] L.A. Isupova, I.P. Prosvirin, Xray photoelectron spectroscopy investigation of perovskites La 1-xSrxFeO3-y (0 ≤ x < 1.0), prepared via a mechanochemical route, Russ. Chem. Bull. 62 (2013) 1564–1569. https://doi.org/10.1007/s11172-013-0225-8.

[10] V. Sadykov, E. Pikalova, N. Eremeev, A. Shubin, I. Zilberberg, I. Prosvirin, E. Sadovskaya, A. Bukhtiyarov, Oxygen transport in Pr nickelates: Elucidation of atomic-scale features, Solid State Ionics. 344 (2020) 115155. https://doi.org/10.1016/j.ssi.2019.115155.

[11] A.I. Stadnichenko, A.M. Sorokin, A.I. Boronin, XPS, UPS, and STM studies of nanostructured CuO films, J. Struct. Chem. 49 (2008) 341–347. https://doi.org/10.1007/s10947-008-0133-1.

[12] G. Kresse, J. Hafner, Ab initio molecular dynamics for liquid metals, Phys. Rev. B. 47 (1993) 558–561. https://doi.org/10.1103/PhysRevB.47.558.

[13] G. Kresse, J. Furthmüller, Efficiency of ab-initio total energy calculations for metals and semiconductors using a plane-wave basis set, Comput. Mater. Sci. 6 (1996) 15–50. https://doi.org/10.1016/0927-0256(96)00008-0.

[14] P.E. Blöchl, Projector augmented-wave method, Phys. Rev. B. 50 (1994) 17953–17979. https://doi.org/10.1103/PhysRevB.50.17953.

[15] G. Kresse, D. Joubert, From ultrasoft pseudopotentials to the projector augmented-wave method, Phys. Rev. B. 59 (1999) 1758–1775. https://doi.org/10.1103/PhysRevB.59.1758.

[16] V.I. Anisimov, J. Zaanen, O.K. Andersen, Band theory and Mott insulators: Hubbard U instead of Stoner I, Phys. Rev. B. 44 (1991) 943–954. https://doi.org/10.1103/PhysRevB.44.943.

[17] S. Dudarev, G. Botton, Electron-energy-loss spectra and the structural stability of nickel oxide: An LSDA+U study, Phys. Rev. B - Condens. Matter Mater. Phys. 57 (1998) 1505–1509. https://doi.org/10.1103/PhysRevB.57.1505.

[18] G. Henkelman, H. Jónsson, A dimer method for finding saddle points on high dimensional potential surfaces using only first derivatives, J. Chem. Phys. 111 (1999) 7010–7022. https://doi.org/10.1063/1.480097.

[19] A. Heyden, A.T. Bell, F.J. Keil, Efficient methods for finding transition states in chemical reactions: Comparison of improved dimer method and partitioned rational function optimization method, J. Chem. Phys. 123 (2005). https://doi.org/10.1063/1.2104507.

[20] A.S. Konopatsky, D. V. Leybo, V. V. Kalinina, I.L. Zilberberg, L.Y. Antipina, P.B. Sorokin, D. V. Shtansky, Synergistic Catalytic Effect of Ag and MgO Nanoparticles Supported on Defective BN Surface in CO Oxidation Reaction, Materials (Basel). 16 (2023). https://doi.org/10.3390/ma16020470.

[21] M. Abbate, G. Zampieri, J. Okamoto, A. Fujimori, S. Kawasaki, M. Takano, X-ray absorption of the negative charge-transfer material (formula presented), Phys. Rev. B - Condens. Matter Mater. Phys. 65 (2002) 1–7.

https://doi.org/10.1103/PhysRevB.65.165120.

[22] T. Takeda, Y. Yamaguchi, H. Watanabe, Magnetic Structure of SrFeO3, J. Phys. Soc. Japan. 33 (1972) 967–969.

[23] W.C. Koehler, E.O. Wollan, Neutron-diffraction study of the magnetic properties of perovskite-like compounds LaBO3, J. Phys. Chem. Solids. 2 (1957) 100–106. https://doi.org/10.1016/0022-3697(57)90095-1.

[24] M. V. Patrakeev, I.A. Leonidov, V.L. Kozhevnikov, K.R. Poeppelmeier, p-type electron transport in La1-xSrxFeO 3-δ at high temperatures, J. Solid State Chem. 178 (2005) 921–927. https://doi.org/10.1016/j.jssc.2004.10.038.